\documentstyle[prl,aps,epsf,twocolumn]{revtex}

\newcommand{\la}{\langle}
\newcommand{\ra}{\rangle}

\newcommand{\beq}{\begin{eqnarray}}
\newcommand{\eeq}{\end{eqnarray}}

\newcommand{\btem}{\bibitem}
\newcommand{\delslash}{\partial \hspace{-6pt}/}

\begin{document}

\draft

\title{In-medium $\pi\pi$ Correlation Induced by \\
 Partial Restoration of Chiral Symmetry}

\author{D. Jido$^{(1)}$, T. Hatsuda$^{(2)}$, T. Kunihiro$^{(3)}$}
\address{$^{(1)}$  Consejo Superior de Investigaciones Cient\'{\i}ficas,
Universitat de Valencia, IFIC, \\
 Institutos de Investigaci\'{o}n de Peterna, Aptdo.
Correos 2085, 46071, Valencia, Spain}
\address{$^{(2)}$  Department of Physics, University of Tokyo,
  Tokyo 113-0077 Japan}
\address{$^{(3)}$ Yukawa Institute for Theoretical Physics,
Kyoto University, Kyoto 606-8502, Japan}

\date{\today}

\maketitle

\begin{abstract}
 We show that  both the linear  and the
 non-linear chiral models give an enhancement of the $\pi\pi$
  cross section near the 2${\pi}$ threshold in the
 scalar-iso-scalar ($I$=$J$=0) channel in nuclear matter.
    The reduction of the chiral condensate, i.e.,
 the partial chiral restoration in nuclear matter,
 is responsible for the enhancement  in both cases.
 We extract an effective 4$\pi$-nucleon vertex which
  is responsible for the enhancement but has not been
   considered in the non-liear models for in-medium $\pi\pi$ interaction.
  Relation of this vertex
   and a next-to-leading order terms in the heavy-baryon chiral
   lagrangian, ${\cal L}_{\pi N}^{(2)}$,
 is also discussed.

\end{abstract}



Exploring possible evidence of partial restoration
 of chiral symmetry  in hot and/or
dense nuclear medium is an intriguing subject in hadron physics
 (see e.g., \cite{HKBR,hirschegg}).
 In particular, the softening of the
 scalar-iso-scalar  ($I=J=0$) fluctuation in matter could be a
 distinct signal of
 the  partial restoration of chiral symmetry, as was originally
 pointed out in \cite{HK85a}.

 In a recent paper \cite{HKS}, Shimizu and two of the present authors
 have shown, using the linear $\sigma$ model (L$\sigma$M),
 that the partial restoration of chiral symmetry in nuclear matter
 leads to an enhanced spectral function
  near the 2${\pi}$ threshold in the $\sigma$-channel.
  (The similar effect in hot
 hadronic plasma has been  considered before \cite{CH,VOL}.)
  They also suggested that
  the enhancement of the  cross section  in the reaction
  $\pi A \rightarrow  \pi^{+}\pi^{-} A'$
  near 2${\pi}$ threshold reported by
 the CHAOS collaboration  \cite{CHAOS}, which was
 originally motivated by a possibility of a strong
 $\pi\pi$ correlation in matter \cite{GSI}, may be
 interpreted as an evidence of the partial restoration of chiral
 symmetry.

 In fact, the state-of-the-art
  calculations using the nonlinear chiral
  lagrangian together with  $\pi N$ many-body
  dynamics do not reproduce the cross sections in the $I=0$ and
  $I=2$ channels simultaneously  \cite{WO}.
  In those calculations,
  the final state interactions of the emitted two pions in nuclei
  give rise to a slight enhancement of the cross section in the
  $I=0$ channel, but is
  not sufficient to reproduce the experimental  data.
 This indicates that
 some additional mechanism such as the partial restoration of chiral
 symmetry  may be relevant for explaining the data.
 Such effect can be readily incorporated
 in the L$\sigma$M as adopted in \cite{HKS,SHUCK}, while
  it is not obvious  how to take into account the effect
 in the non-linear chiral models.

The purposes of the present paper are twofold: Firstly,
 we show that the near-threshold enhancement in the $I=J=0$ channel
in nuclear matter takes place
 irrespective of the representations (linear or non-linear) of
 chiral symmetry, if
 the effect of partial restoration of chiral symmetry
in the medium is  properly incorporated.
 Secondly, we show that ${\cal L}_{\pi N}^{(2)}$,
  which is a next-to-leading order term in the
   non-linear chiral lagrangian in the heavy-baryon formalism,
 gives relevant vertices for the  near-threshold enhancement
 in the $I$=$J$=0 channel.
  Such vertices have not been taken into account in
  previous calculations of the in-medium $\pi\pi$ scattering
   in the non-linear lagrangian approaches.

 Let us first consider the $\pi\pi$ scattering in nuclear matter
 in the  SU(2) L$\sigma$M
 to  demonstrate the essential idea of the near-threshold enhancement.
 The following argument is slightly different from the
 original one \cite{HKS} but is
 easy to be generalized to the non-linear case.
 The lagrangian adopted in \cite{HKS} is the standard Gell-Mann-Levy
 model with a minor modification
\beq
\label{model-l}
{\cal L} & = &  {1 \over 4} {\rm Tr} [\partial M \partial M^{\dagger}
 - \mu^2 M M^{\dagger} - {2 \lambda \over 4! } (M M^{\dagger})^2
 \nonumber \\
 & - & h (M+M^{\dagger}) ]
  + \bar{\psi} ( i \delslash - g M_5 ) \psi
  + \cdot \cdot \cdot  ,
\eeq
where  $M = \sigma + i \vec{\tau}\cdot \vec{\pi}$,
 $M_5 = \sigma + i \gamma_5 \vec{\tau}\cdot \vec{\pi}$,
 $\psi$ is the nucleon field, and
 Tr is for the flavor index \cite{add}.
 In this model, the difference of the
 $\pi\pi$ scattering in the vacuum and that in nuclear matter
 stems from  the nucleon-loops which modify the
 self-energies and vertices of $\pi$ and $\sigma$.
 The effect of the
 nucleons in nuclear matter on the low-energy pions are suppressed
 due to chiral symmetry.
 Therefore,  one-loop contributions leading to the substantial
 modifications of the in-medium $\pi\pi$
 scattering come  mainly from the diagrams shown in Fig.1.
  Here, Fig.1(a) and Fig.1(b)
  are the medium modifications of the  $\sigma$ propagator
  and the $\sigma\pi\pi$ vertex, respectively.
  Those  mean-field  contributions
  can be
 incorporated by making the following replacement
 in the $\pi\pi$ amplitude in the vacuum;
\beq
\la \sigma \ra_0 \rightarrow \la \sigma \ra
 =  \ \Phi (\rho) \la \sigma \ra_0  ,
\eeq
where $\la \cdot \ra_0$ ($\la \cdot \ra$) denotes the
 vacuum (nuclear matter) expectation value, and
 $\rho$ ($\rho_0$) is the baryon density (nuclear matter
 saturation density).
 $\Phi(\rho)$ may be parameterized as
 $\Phi(\rho) = 1 - C \rho / \rho_0 $ with
 $C = 0.1 -0.3$ \cite{HKS}.

The in-medium $\pi\pi$ scattering amplitude in the tree level
 is  written as $T_{\rm tree}(s,t,u)
 =  \delta^{ab} \delta^{cd}
 A(s,t,u) + ({\rm crossings})$
with
\beq
\label{scatt-a}
A(s,t,u) \equiv A(s) =
 - {\lambda \over 3}\ {s-m_{\pi}^2 \over s-m_{\sigma}^{*2}},
\eeq
where  $m_{\sigma}^*$ is the in-medium $\sigma$ mass defined by
$m_{\sigma}^{*2} = \lambda \la \sigma \ra^2/3 + m_{\pi}^{2}$.
The scattering amplitude after the projection to
 $I=J=0$ channel reads
\beq
\label{ampI=0}
T_{\rm tree} (s) & = & {1 \over 2} \int_{-1}^{+1} d(\cos\theta )
 \ [ 3 A(s) + A(t) + A(u) ] \ .
\eeq
The tree-level amplitude generally breaks unitarity
  because the imaginary part is not incorporated.
 A simplest way to restore unitarity is to
 start with a trivial identity for the full amplitude,
  $T(s)= ({\rm Re}T^{-1} + i {\rm Im}T^{-1})^{-1}$, and
 to evaluate the inverse amplitude
 $T^{-1}$ rather than $T$  \cite{OOP}.
  For the $I=J=0$ scattering below the
 4$\pi$ threshold, the real part of $T^{-1}$
 may be approximated by the tree amplitude,
 ${\rm Re}T^{-1} \simeq T^{-1}_{\rm tree}$, while
 the unitarity condition $ i (T^{\dagger} - T) = T^{\dagger}T $
  determines the imaginary part as
${\rm Im}T^{-1} = - \Theta(s) /2 =
 -  \theta(s - 4 m_{\pi}^2) {1 \over 32
\pi} \sqrt{1 - {4m_\pi^2 / s}} $.  Here
 $\Theta(s)$ is  the phase-space volume of the
 2$\pi$ intermediate state.
 Thus one arrives at the unitarized amplitude
\begin{equation}
   \label{newT}
 T(s) = {1  \over  T_{\rm tree}^{-1}(s) - i \Theta(s)/2} .
\end{equation}

  The in-medium $\pi\pi$ cross section is obtained by multiplying the
 flux factor:
$\sigma_{\pi\pi}(s;\rho) \propto  | T(s) |^2 /s$.
Shown in Fig.2 is the cross section
 for the $I=J=0$ channel in arbitrary unit.
  One can see the enhancement near 2$\pi$ threshold
associated with the decrease of the chiral condensate $\la \sigma \ra$.
The parameters in the lagrangian are chosen so that
$m_\pi$=140MeV,
  $\la \sigma \ra_0$=$f_{\pi}$=93MeV and
 $m_{\sigma}^*(\rho=0)$=550MeV (which corresponds to
 $\lambda/4\pi= 7.8$) are reproduced.
 We have checked that the case for
 $m_{\sigma}^*(\rho=0)$=750MeV ($\lambda /4\pi= 15.0$)
 does not show qualitative change.
 The scattering lengths for $(I=0, J=0)$ and $(I=2, J=0)$ channels
   are well reproduced in these parameters.
   In Fig.3,
the in-medium cross section
 relative to its vacuum value defined below is shown;
\beq
\label{R-ratio}
R={ \sigma_{\pi\pi}(s;\rho)
  \over \sigma_{\pi\pi}(s;\rho=0)}.
\eeq
 As is anticipated, a  large enhancement near 2${\pi}$ threshold
 can be seen in Fig. 3.
 However, to make a realistic comparison of our schematic  calculation
 with the experimental data\cite{CHAOS},
 one needs to incorporate the amplitudes describing the elementary
processes of the pion production and absorption
in the reaction $\pi A \rightarrow  \pi \pi A'$. This  is
 beyond the scope of this work.

The near-threshold enhancement of $T(s)$ and $R$
in eqs. (\ref{newT},\ref{R-ratio})
takes place in association with partial restoration of chiral symmetry
in nuclear matter. 
In fact, as  $\la \sigma \ra$ decreases,
$m_{\sigma}^*$   approaches
2$m_{\pi}$, which implies that $T_{\rm tree}^{-1}(s \simeq 2 m_{\pi})$
 tends to be  suppressed. Hence the $s$-dependence of
 the full inverse amplitude $T^{-1}(s)$ just above the 2$\pi$ threshold
 is governed 
 by the imaginary part $\Theta(s  ) /2$, which causes the
  near-threshold enhancement of $T(s)$.
This is consistent with the observation made in
 Ref.\cite{HKS}, where the spectral function
 in the $\sigma$-channel rather than the $T$-matrix is shown to have
 a near-threshold enhancement by the same mechanism \cite{note}.

 At this point, a natural question to ask is
 whether the near-threshold enhancement obtained in L$\sigma$M
 arises also in the non-linear models.
 Furthermore, if it is the case,
 what kind of vertices in the non-linear chiral lagrangian
 are responsible for the enhancement?
 To study these problems,
 we start with the standard polar parameterization of the chiral field,
 $M = \sigma + i \vec{\tau} \cdot \vec{\pi}
  = (\la \sigma \ra + S) U $ with $U = \exp (i \vec{\tau}
 \cdot \vec{\phi} /f^{*}_{\pi})$. Here,
$\la \sigma \ra$ is the sigma condensate in nuclear matter
 as before, while
 $f^{*}_{\pi}$ is a would-be ``in-medium pion decay constant''
  taken as
an arbitrary constant for the moment.
 Putting the parameterization into eq.(\ref{model-l}) and making
 suitable redefinition of the nucleon field from $\psi$ to $N$,
 one obtains the non-linear form of (\ref{model-l}),
\beq
\label{model-nl}
{\cal L} & = &  {1 \over 2} [(\partial S)^2 - m_{\sigma}^{*2} S^2]
  - {\lambda \la \sigma \ra \over 6} S^3 - {\lambda \over 4!} S^4 \nonumber
\\
 & + & {(\la \sigma \ra +S)^2 \over 4} {\rm Tr}
 [\partial U \partial U^{\dagger}] + { \la \sigma \ra + S \over 4}\  h \
 {\rm Tr}[U^{\dagger}+U] \nonumber \\
& + & {\cal L}_{\pi N}^{(1)} - g S \bar{N} N \ ,
\eeq
with
\beq
{\cal L}_{\pi N}^{(1)} =
\bar{N}(i \delslash + i v \hspace{-6pt}/
 + i  a \hspace{-6pt}/   \gamma_{5}  - m_{N}^* ) N
\eeq
where $(v_{\mu},a_{\mu}) = (\xi \partial_{\mu} \xi^{\dagger}
 \pm \xi^{\dagger} \partial_\mu \xi)/2$, and
 $m_N^* = g \la \sigma \ra$.
  $\la \sigma \ra$ is to be determined by minimizing the
 effective potential in nuclear matter calculated with
eq.(\ref{model-nl}), or equivalently by the condition
 $\la S \ra =0$.
 In the mean-field level, the density dependence of
  $\la \sigma \ra$ is dictated solely by
 the Yukawa coupling, $S \bar{N}N$.

 It is worth emphasizing that $f^{*}_{\pi}$,
 which has been left undetermined,
 should be chosen as $f^{*}_{\pi}=\la \sigma \ra$
  to impose proper normalization of the pion field
 in nuclear matter. This is in fact   crucial for
 the  linear and non-linear representations to give
 equivalent results  in nuclear matter.

 In-medium $\pi\pi$ amplitude $A(s)$ in the tree level
 obtained from (\ref{model-nl}) reads
 \beq
\label{scatt2}
A(s) =  {s- m^2_\pi \over \la \sigma \ra^2}
       - {(s - m_\pi^2)^2 \over
 \la \sigma \ra^2} {1 \over s - m_{\sigma}^{*2}}  .
\eeq
The first term in (\ref{scatt2}) comes from the
 contact 4$\pi$ coupling generated by  the
 expansion of the second line in (\ref{model-nl}) with
 the coefficient proportional to
 $1 / {\la \sigma \ra}^2$.
 On the other hand,
 the second term in (\ref{scatt2}) is from the contribution of the
  scalar meson $S$ in the $s$-channel.

Of course, $A(s)$ in eq.(\ref{scatt2}) and  its unitarized
 amplitude  $T(s)$  are exactly the same with
 eq. (\ref{scatt-a}) and eq.(\ref{newT}), respectively,
 since what we have done is simply the field redefinition.
 Nevertheless, we have  two observations from the {\em chiral
   invariant} decomposition eq.(\ref{scatt2}):

\noindent
(i) If $m_{\sigma}^* $ ($= \lambda \la \sigma \ra^2/3 + m_{\pi}^2$)
  is finite and decreases as
 density increases, the near-threshold enhancement occurs exactly in
 the same way as the previous case shown in Fig.2 and Fig.3.

\noindent
(ii) Even in the limit where
  $m_{\sigma}^*$ is infinitely large ($\lambda \rightarrow \infty$)
  and hence the second term of eq.(\ref{scatt2}) is negligible,
  the first term  signals the  enhanced
  attraction of the $\pi\pi$ interaction
  as long as $\la \sigma \ra $ decreases in nuclear matter.

 The second observation
  is demonstrated in Fig.4 where
 the in-medium $\pi\pi$ cross section in the $I$=$J$=0 channel
  is calculated with the unitarized amplitude obtained only from
  the first term of (\ref{scatt2}).
 Although the enhancement is smaller than that of
 L$\sigma$M, there is still a large enhancement in the
 narrow range of the $\pi\pi$ invariant mass near 2${\pi}$ threshold.
  One should note here that as $\la \sigma \ra$ decreases,
  the convergence of  the chiral expansion of $A(s)$ by $s/\la \sigma \ra$
  becomes less reliable. Therefore, Fig.4 with only the
   leading term in the chiral expansion should be taken as a
    qualitative one.

It is in order here to emphasize
 the relation of the near-threshold enhancement and the
 softening of the $I=J=0$ fluctuation in
 nuclear matter. Such fluctuation can be characterized by the
 complex pole of the unitarized amplitude $T(s)$  in eq.(\ref{newT}).
 For simplicity, let us take the case (ii) in
  the chiral limit, where  the analytic solution can be written as
 $\sqrt{s_{\rm pole}} = \sqrt{8\pi} \la \sigma \ra \ (1-i) $.
 This complex pole corresponds to a broad $I=J=0$  mode dynamically
 generated by the $\pi\pi$ rescattering.
  As $\la \sigma \ra$ is reduced in the medium, the pole
 moves toward the origin in the complex
 $\sqrt{s}$-plane \cite{Oller} and thus causes an enhancement in the
 $\pi\pi$ cross section as shown in Fig.4.
 It is also true for the case (i) leading to Fig.2.
  Such connection between the moving pole in the complex plane
 and the near-threshold enhancement
 has been  first pointed out in \cite{HK85b} using
 the Nambu-Jona-Lasinio model.
 Further  study in the present context with finite $m_{\pi}$
 will be discussed elsewhere \cite{JHK2}.

 Now, we turn to the question on the effective
  vertex  responsible for the
 enhanced $\pi\pi$ attraction in nuclear matter
 in the non-linear chiral lagrangian.
 To study this, let us start with
  eq. (\ref{model-nl})  in the {\em vacuum} and
 take a heavy-scalar ($S$) limit and heavy-baryon ($N$) limit
 simultaneously.
 These limits can be achieved by
 $\lambda, g \rightarrow \infty$ with
 $g/\lambda$ and $\la \sigma \ra_0 = f_{\pi}$ being fixed.
 In this limit,
 the heavy scalar field  $S$ may be integrated out
 and the following effective lagrangian is obtained:
\beq
\label{model-nl2}
{\cal L} & = &
 \left(
 {f_{\pi}^2 \over 4}
 - {g f_{\pi}  \over 2 m_{\sigma}^2}\bar{N}N
 \right)
\left(
{\rm Tr} [\partial U \partial U^{\dagger}]
 - {h \over f_{\pi}} \  {\rm Tr}[U^{\dagger}+U] \right)
 \nonumber \\
 &  & \ + \  {\cal L}_{\pi N}^{(1)} + \cdot \cdot \cdot \ \ ,
\eeq
where all the constants take their vacuum values:
 $f_{\pi} = \la \sigma \ra_0 $,
 $m_{\sigma}^2 = \lambda \la \sigma \ra_0^2/3 + m_{\pi}^2$,
 and  $m_N = g \la \sigma \ra_0$.
 Note that $g f_{\pi} / 2  m_{\sigma}^2$
 in front of $\bar{N}N$ approaches to
 a finite value
 $3 g/2\lambda f_{\pi}$ in the heavy limit, thus it cannot be neglected.
 In (\ref{model-nl2}),  $\cdot \cdot \cdot$
  denotes
  other higher dimensional operators which are not relevant for the
   discussion below.

 In the uniform nuclear matter, $\bar{N}N$ in
 eq.(\ref{model-nl2}) may be replaced by $\rho$.
 This leads to a reduction of  the vacuum condensate;
 $f_{\pi} = \la \sigma \ra_0  \rightarrow
  \la \sigma \ra =
 \la \sigma \ra_0 (1- g  \rho/f_{\pi} m_{\sigma}^2)
 = f_{\pi}^*$.
Then the proper normalization of the
 pion field  in nuclear matter should be
 $\phi' = (\phi /f_{\pi}) \cdot f_{\pi}^* $.
 The in-medium $\pi\pi$ scattering
 with this normalization exactly reproduces
  the first term in (\ref{scatt2}) as it should be.
 Therefore, the origin of the
 near-threshold enhancement in the heavy  limit
 can be ascribed to the following  new vertex:
\beq
\label{new-vertex}
{\cal L}_{\rm new} = - {3g \over 2 \lambda f_{\pi}} \
\bar{N}N {\rm Tr} [\partial U \partial U^{\dagger}].
\eeq
Because this vertex is proportional to
 the scalar-isoscalar density of the nucleon,
  it affects not only the pion propagator  but also
 the interaction among pions in nuclear matter.
 In Fig.5,  4$\pi$-$N$-$N$ vertex generated by ${\cal L}_{\rm new}$
 is shown as an example.
 Note that the vertex in Fig.5 acts to enhance the $\pi\pi$ attraction in
 the $I=J=0$ channel in nuclear matter. Despite its important role,
 this vertex has not been considered so far  in the calculations
of  the $\pi\pi$ scattering amplitudes in nuclear matter in the
 non-linear approaches.

 Here it should be pointed out that
 the vertex eq.(\ref{new-vertex})
 which  we have extracted from the L$\sigma$M
 has been known to be one of the next-to-leading order terms
 in the non-linear chiral lagrangian in the
  heavy-baryon formalism  \cite{GSS,BKM}.
 In fact, the chiral lagrangian with the nucleon
 up to ${\cal O}(Q^2)$ of the chiral counting reads \cite{GSS,BKM}
\beq
\label{212}
{\cal L} = {\cal L}_{\pi \pi}^{(2)}
+ {\cal L}_{\pi N}^{(1)} + {\cal L}_{\pi N}^{(2)},
\eeq
where $ {\cal L}_{\pi \pi}^{(2)}$ is the standard
 ${\cal O}(Q^2)$ lowest order chiral lagrangian  for the pion,
  $ {\cal L}_{\pi N}^{(1)}$ (essentially the same as
 $ {\cal L}_{\pi N}^{(1)}$
 in (\ref{model-nl})) is the ${\cal O}(Q)$
  term with a single derivative.
 $ {\cal L}_{\pi N}^{(2)}$ is an ${\cal O}(Q^2)$ term
 having the following structure
  (the complete structure is given in Appendix A of \cite{BKM}),
\beq
\label{gss}
 {\cal L}_{\pi N}^{(2)} & = &  c_3 \bar{N} (u \cdot u) N
 + (c_2 - {g_A^2 \over 8 m_N})  \bar{N}({\rm v} \cdot u)^2 N \nonumber \\
 &  & +  c_1 \bar{N}N {\rm Tr} (U^{\dagger}  \chi + \chi^{\dagger} U)
      + \cdot \cdot \cdot ,
\eeq
where $u_{\mu} = i \xi^{\dagger} \partial_{\mu} U \xi^{\dagger}$,
$U = \xi^2 = {\rm exp}(i \tau^a \phi^a /f_{\pi})$,
 ${\rm v}_{\mu}$ is the four velocity of the nucleon,
 $g_A$ is the axial charge of the nucleon, and
 $\chi =  h/f_{\pi}$ denotes the explicit symmetry breaking.
 Thorsson and Wirzba \cite{TW} have derived the in-medium
 chiral lagrangian from eq.(\ref{gss}) by taking the mean-field
approximation
 for the nucleon field;
\beq
\label{mean-f}
\la {\cal L} \ra
 & = & ( {f_{\pi}^2 \over 4} + {c_3 \over 2} \rho)\  {\rm Tr}
 [\partial U \partial U^{\dagger}] \nonumber \\
 &  &  + \ \  ( {c_2 \over 2} - {g_A^2 \over 16 m_N})\ \rho \  {\rm Tr}
 [\partial_0 U \partial_0 U^{\dagger}] \nonumber \\
 &  & +  \ \
  ({ f_{\pi}^2 \over 4} + {c_1 \over 2} \rho)
 \ {\rm Tr} (U^{\dagger}  \chi + \chi^{\dagger} U).
\eeq
Note that ${\cal L}_{\pi N}^{(1)}$ in (\ref{212})
 disappears in (\ref{mean-f}) because of the isospin symmetry
 in nuclear matter.

  As is shown in \cite{BKM}, the coefficients $c_{1,2,3}$ are related
 to the $\pi N$ sigma term, the
 iso-spin even
 $S$-wave $\pi N$ scattering length, and
 the nucleon axial polarizability, respectively.
 We just quote here the numbers although they  have large potential
 uncertainties:
 $c_1 = -0.87 {\rm GeV}^{-1}$,  $c_2 = 3.34 {\rm GeV}^{-1}$,
 and  $c_3 = -5.25 {\rm GeV}^{-1}$.

 By comparing eq.(\ref{model-nl2}) with
 eq.(\ref{gss}) or eq.(\ref{mean-f}), we observe that
 the new vertex of the form
 $\bar{N}N {\rm Tr} [\partial U \partial U^{\dagger}]$
 appear in both cases with the same sign.
 In  eq.(\ref{model-nl2}), we have only the Lorentz invariant terms
 because we have integrated out only the scalar meson $S$.
 On the other hand,
 eq.(\ref{mean-f}) contains more general $O(3)$ invariant
 terms because it potentially contains the effect of
 heavy excited baryons \cite{BKM}.

 Finally, let us comment on the recent analysis by Oset and
Vicente Vacas on the
 $\pi\pi$ scattering in nuclear matter using non-linear
 chiral lagrangian \cite{OV}.
 They adopted only
  ${\cal L}_{\pi \pi}^{(2)}+{\cal L}_{\pi N}^{(1)}$ for the
 $\pi N$ dynamics and concluded that
 the near-threshold enhancement is not obtained from the
 4$\pi$-$N$-$N$ vertex generated by ${\cal L}_{\pi N}^{(1)}$.
 This observation is  consistent with what we have mentioned
  before; the effect of
  ${\cal L}_{\pi N}^{(1)}$ disappears due to isospin symmetry
 and has no significance
 on the threshold $\pi\pi$ correlation.
  Therefore, it is of
  of great importance to make
 an extensive analysis of the $\pi \pi$ interaction in nuclear matter
 with  ${\cal L}_{\pi N}^{(2)}$.

In summary, we have shown that the near-threshold enhancement of the
in-medium
 $\pi\pi$ scattering in the $I=J=0$ channel occurs
 both in the linear and nonlinear representation of chiral symmetry.
 The enhancement is caused by the
  in-medium reduction of the  chiral condensate
 associated with partial restoration of chiral symmetry.
 In fact,  the decrease of the chiral condensate is shown to
 enhance the $\pi\pi$ attraction in the $I=J=0$ channel and simultaneously
 induce
 the  softening of the scalar-iso-scalar fluctuations in nuclear matter.
 Also, we have identified an effective 4$\pi$-nucleon vertex
 responsible for the enhancement and discussed its relation to
  the   next-to-leading order term  ${\cal L}_{\pi N}^{(2)}$
  in the non-linear chiral lagrangian in the heavy-baryon formalism.
  It is thus an intriguing problem to examine
 whether  the softening of the scalar-iso-scalar fluctuation
 induced by ${\cal L}_{\pi N}^{(2)}$ manifests itself
  in explaining the CHAOS data in the non-linear lagrangian
 approach. 

T.H. Thanks J. Wambach,  G. Chanfray, P. Schuck, E. Oset, M. J. Vicente
Vacas
 and N. Grion for useful discussions.
This work was partially supported by the Grants-in-Aid of
the Japanese Ministry of Education, Science and Culture
(No. 12640263 and 12640296).
The work by D.J. was supported by the Spanish Ministry of Education in the
Program
``Estancias de Doctores y Tecn\'ologos Extranjeros en Espa\~{n}a''.


\begin{figure}[tbp]
   \centering
   \epsfxsize=4.0cm
   \epsfbox{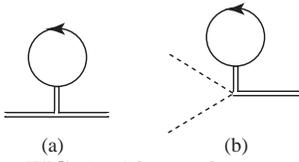}
    \caption{The medium modifications of (a) the $\sigma$ propagator
    and (b) the $\sigma \pi\pi$ vertex through the nucleon-loop
 in the mean-field approximation. The solid line with arrow,
 the dashed line and the double line represent the nucleon, $\pi$ and
 $\sigma$, respectively.}
    \label{fig1}
\end{figure}

\begin{figure}[tbp]
   \centering
   \epsfxsize=8.0cm
   \epsfbox{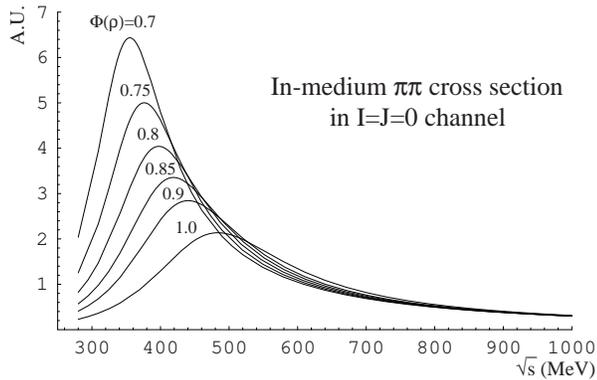}
    \caption{In-medium $\pi\pi$ cross section in the $I=J=0$ channel
 for different values of $\Phi(\rho)$.
    The cross section is shown in the arbitrary unit (A.U.). }
    \label{fig2}
\end{figure}

\begin{figure}[tbp]
   \centering
   \epsfxsize=8.0cm
   \epsfbox{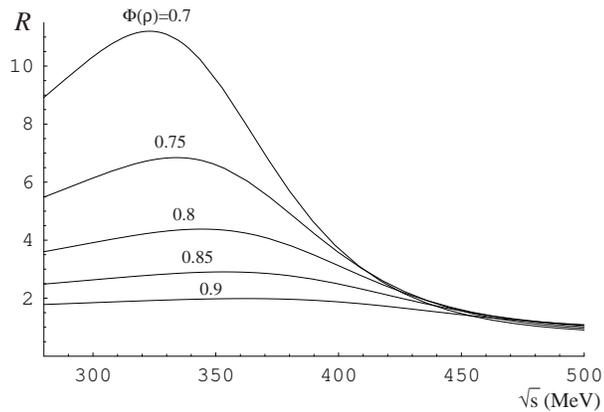}
    \caption{$R$ (the in-medium cross section divided its vacuum value)
 in the $I=J=0$ channel.}
    \label{fig3}
\end{figure}

\begin{figure}[tbp]
   \centering
   \epsfxsize=8.0cm
   \epsfbox{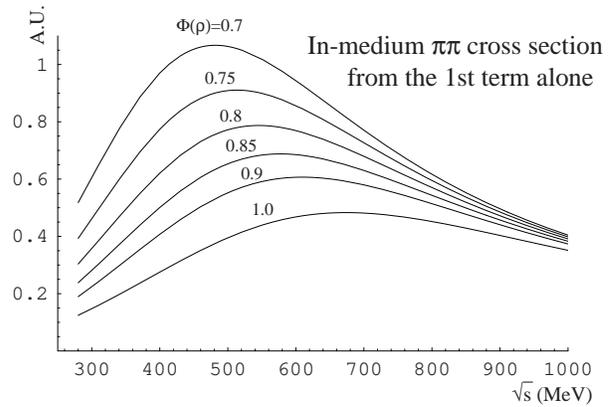}
    \caption{In-medium $\pi\pi$ cross section in the $I=J=0$ channel
 in the heavy $S$ limit where $m_{\sigma}^*$ is taken to be infinity.
    The cross section is shown in the arbitrary unit (A.U.) but
 in the same scale with Fig.2.}
    \label{fig4}
\end{figure}

\begin{figure}[tbp]
   \centering
   \epsfxsize=2.5cm
   \epsfbox{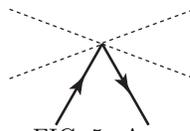}
    \caption{An example of the new 4$\pi$-$N$-$N$ vertex generated
 by ${\cal L}_{\rm new}$ or by ${\cal L}_{\pi N}^{(2)}$.
 The solid line with arrow and the dashed line represent
 the nucleon and pion, respectively.}
    \label{fig5}
\end{figure}

\end{document}